\documentclass[letter]{aa}
\usepackage{graphicx}
\usepackage{txfonts}
\usepackage{hyperref}
\usepackage{natbib}
\usepackage{amsmath}
\usepackage{amssymb}
\usepackage{placeins}
\usepackage{xcolor}

\begin{document}

    \title{The dispersion of $E_{\rm p,i}$--$L_{\rm iso}$ correlation of long gamma--ray bursts is partially due to assembling different sources}

    \titlerunning{Time-resolved $E_{\rm p,i}$--$L_{\rm iso}$ within individual GRBs}

    \author{M.~Maistrello\thanks{mstmnl@unife.it}\inst{\ref{inst1}}
        \and R.~Maccary\inst{\ref{inst1}}
        \and C.~Guidorzi\inst{\ref{inst1},\ref{inst2},\ref{inst3}}
        \and L.~Amati\inst{\ref{inst3}}
        }
            
    \institute{Department of Physics and Earth Science, University of Ferrara, Via Saragat 1, I-44122 Ferrara, Italy
            \label{inst1}
        \and 
            INFN –- Sezione di Ferrara, Via Saragat 1, 44122 Ferrara, Italy 
            \label{inst2}
        \and 
            INAF –- Osservatorio di Astrofisica e Scienza dello Spazio di Bologna, Via Piero Gobetti 101, 40129 Bologna, Italy
            \label{inst3}
            }
    
    \date{Received date / Accepted date }
    
    \abstract
    {Long gamma-ray burst (GRB) prompt emission shows a correlation between the intrinsic peak energy, $E_{\mathrm{p,i}}$, of the time-average $\nu F_{\nu}$ spectrum and the isotropic-equivalent peak gamma-ray luminosity, $L_{{\rm p,iso}}$, as well as the total released energy, $E_{\rm iso}$. The same correlation is found within individual bursts, when time-resolved $E_{\rm p,i}$ and $L_{\rm iso}$ are considered. These correlations are characterised by an intrinsic dispersion, whose origin is still unknown. Discovering the origin of the correlation and of its dispersion would shed light on the still poorly understood prompt emission and would propel GRBs to powerful standard candles.}
    {We studied the dispersion of both isotropic-equivalent and collimation-corrected time-resolved correlations. We also investigated whether the intrinsic dispersion computed within individual GRBs is different from that obtained including different bursts into a unique sample. We then searched for correlations between key features, like Lorentz factor and jet opening angle, and intrinsic dispersion, when the latter is treated as one of the characterising properties.}
    {We performed a time-resolved spectral analysis of 20 long Type-II or collapsar-candidate GRBs detected by the Fermi Gamma-ray Burst Monitor with known redshift and estimates of jet opening angle and/or Lorentz factor. Time intervals were determined using Bayesian Blocks. Then we carried out a statistical analysis starting from distributions of simulated values of the intrinsic dispersion of each burst in the sample.}
    {The collimation-corrected correlation appears to be no less dispersed than the isotropic-equivalent one. Also, individual GRBs are significantly less dispersed than the whole sample. We excluded (at $4.2 \sigma$ confidence level) the difference in samples' sizes as the possible reason, thus confirming that individual GRBs are {\em intrinsically} less dispersed than the whole sample. No correlation was found between intrinsic dispersion and other key properties for the few GRBs with available information.}
    {The contribution to the dispersion by the jet opening angle is not relevant. Moreover, our results prove that the intrinsic dispersion which affects the $E_{\rm p,i}$--$L_{{\rm iso}}$ correlation is partially, but not entirely due to assembling different GRBs. We therefore conclude that the presence of different GRBs significantly contributes to the observed dispersion of both time-average $E_{\rm p,i}$--$L_{\rm p,iso}$ and $E_{\rm p,i}$--$E_{\rm iso}$ correlations.}

    \maketitle

%__________________________________________________________________

    \section{Introduction}

        Despite the prompt emission of gamma-ray bursts (GRBs) was discovered more than 50 years ago, several aspects are still poorly understood. One of them concerns some correlations that emerge when considering time-integrated spectra of type-II or collapsar GRBs with known redshift. Specifically, it was found that the intrinsic (that is, redshift-corrected) peak energy, $E_{\rm p,i}$, of the $\nu F_{\nu}$ spectrum correlates with the isotropic-equivalent gamma-ray radiated energy, $E_{\rm iso}$ \citep{Amati02}, with peak luminosity, $L_{\mathrm{p, iso}}$ \citep{Yonetoku04}, and with collimation-corrected energy, $E_{\gamma}$ \citep{Ghirlanda04}. These correlations have extensively been investigated over the last two decades mainly for two reasons: 1) they are key to gain clues on the radiative process(es) and on the dissipation mechanism at play (see \citealt{Kumar15} for a review); 2) they can be exploited to use GRBs as probes of the cosmological parameters (see \citealt{Moresco22} for a review). 
        
        In this work, we focus on the time-resolved $E_{\rm p,i} - L_{\rm iso}$ correlation, which can be modelled as a power--law (PL):
        \begin{equation}
            \label{eq::amatirelation}
            \log \left ( \dfrac{E_{\rm p,i}}{\mathrm{keV}} \right ) = m \, \log \left ( \dfrac{L_{\rm iso}}{10^{52} \, \mathrm{erg} \, \mathrm{s}^{-1}} \right ) + q \, ,
        \end{equation}
        where $m$ and $q$ are the PL index and normalisation, respectively. The correlation is also characterised by an intrinsic dispersion, $\sigma_{\rm int}$, whose origin is still unknown. 
        
        Even if this correlation was discovered starting from time-average spectra of different GRBs, several authors \citep{Ghirlanda10, Frontera12, Lu12, Basak13} found that it holds within individual GRBs too, with a slope and normalisation consistent with the time-integrated values. That the dispersion does not vanish within individual bursts is evidence that its origin cannot be entirely ascribed to properties that differ for each GRB, such as the jet opening angle, as originally suggested by \citet{Ghirlanda04}. Yet, the question remains as to which extent the dispersion must be ascribed to the dissipation mechanism operating within individual bursts and what is, instead, due to assembling different sources. To gain clues, we decided to compare the values of the dispersion of individual bursts with the dispersion of a whole sample of bursts. 
        
        Recently, \citet{Camisasca23a} confirmed that the minimum variability timescale (MVT) of type-II GRBs correlates with peak luminosity and Lorentz factor and found evidence that it may also correlate with the jet opening angle. A possible interpretation builds on 3D general-relativity-magneto-hydrodynamic state-of-the-art simulations of structured, relativistic jet propagating through stellar envelopes and possibly wobbling around the line of sight \citep{Gottlieb22}. Motivated by these results, we tested the possibility that the intrinsic dispersion of individual bursts might be due to the putative wobbling of the jet: in fact, in that case a correlation between $\sigma_{\rm int}$, the jet angle, and/or the Lorentz factor could be expected, in principle.
        
        As a matter of fact, type-I or binary merger candidate GRBs were initially found to be outliers from the $E_{\rm p,i}$--$L_{\rm iso}$ correlation of type-II GRBs. However, it was later found that these GRBs also obey a similar correlation, which is just shifted upward in the $E_{\rm p,i}$--$L_{\rm iso}$ plane (see \citealt{MinaevPozanenko20} and references therein). The different location of type-I GRBs in this plane offers an independent clue to classify GRBs, which can be particularly useful when the light curve (LC) alone is ambiguous (see \citealt{Rossi22, Rastinejad22, Troja22, Yang22, Camisasca23b}). 
        
        In this work we carried out a time-resolved spectral analysis of a sample of 20 type-II GRBs with known redshift, which were detected by the Gamma-ray Burst Monitor (GBM; \citealt{Meegan09}) aboard the Fermi satellite. In Section~\ref{sec::dataanalysis} we describe the sample selection and how the analysis was performed. Results and conclusions are reported in Sections~\ref{sec::results} and \ref{sec::conclusions}, respectively. $\Lambda$CDM cosmology as in \citet{Planck20} was assumed.
        
%__________________________________________________________________

    \section{Sample selection and data analysis}
        \label{sec::dataanalysis}
          
        We used a sample of 20 type-II GRBs detected by Fermi-GBM for which estimates of the redshift $z$, the jet half-opening angle $\theta_{\rm j}$ (both in the homogeneous interstellar medium, ISM, and wind profile, W), and/or the Lorentz factor $\Gamma_0$ were available from the literature. In particular, we took $z$ and $\theta_{\rm j}$ form \citet{Zhao20}, while $\Gamma_0$ from \citet{Ghirlanda18}. Exceptions were done for 171010A and 221010A. For the former, the uncertainties on $\theta_{\rm j}$ reported by \citet{Zhao20} were larger than the estimate of the angle itself, therefore we used the value derived by \citet{Chand19}. For 221010A we took $z$ and $\theta_{\rm j}$ from \citet{Zhu23}. We found no estimate of $\Gamma_0$. The selected bursts are reported in Table~\ref{table::intrinsicdispersion}.
        Hereafter this will be referred to as the $\mathcal{S}$ sample. The subsets of GRBs with estimates of $\theta_{\rm j}^{\rm (ISM)}$ and $\theta_{\rm j}^{\rm (W)}$ will be referred to as the $\mathcal{S}^{\rm (ISM)}$ and $\mathcal{S}^{\rm (W)}$ samples, respectively. The number of bursts in each sample is reported in Table~\ref{table::amaticorrelationparameters}.

        \begin{table*}
            \caption{Sample of GRBs with redshift $z$, jet half-opening angle $\theta_{\rm j}$ (both in the homogeneous interstellar medium, ISM, and wind profile, W), Lorentz factor $\Gamma_0$, and minimum variability timescale (MVT). Values of $z$ and $\theta_{\rm j}$ are from \citet{Zhao20}, $\Gamma_0$ from \citet{Ghirlanda18}, and MVT from \citet{Camisasca23a}. Also reported are best-fit parameters of the $E_{\rm p,i}$--$L_{\rm iso}$ correlation (90\% confidence), number of spectra $N$, and the representative value of $\log E_{\rm p,i}$ computed at a common reference luminosity (see Section~\ref{subsec::searchforcorrelations}).} 
            \label{table::intrinsicdispersion} 
            \centering 
            \begin{tabular}{c c c c c c c c c c c}
            \hline\hline
            GRB     & $z$    & $\theta_{\rm j}^{\mathrm{(ISM)}}$ & $\theta_{\rm j}^{\mathrm{(W)}}$ & $\Gamma_0$        & MVT                       & $m$                    & $q$                    & $\sigma_{\rm int}$ & $N$ & $(\log{(E_{\rm p,i}/{\rm keV}))_{\mathrm{c}}}$ \\ 
                    &        & ($10^{-2}$~rad)                        & ($10^{-2}$~rad)                      &                   & (s)                       &                        &                        &                       &               &                                       \\
            \hline
            090102  & 1.547  & n.a.                                   & n.a.                                 & $215^{+11}_{-10}$ & $0.123_{-0.032}^{+0.043}$ & $0.90_{-0.18}^{+0.18}$ & $-42_{-9}^{+9}$ & $0.16_{-0.04}^{+0.10}$ & 18 & $2.52 \pm 0.08$ \\ 
            
            090323  & 3.5832 & $10.3 \pm 4.3$                         & $4.5 \pm 1.2$                        & $489^{+30}_{-30}$ & $0.465_{-0.120}^{+0.161}$ & $0.44_{-0.24}^{+0.26}$ & $-25_{-11}^{+11}$ & $0.28_{-0.04}^{+0.10}$ & 32 & $2.69 \pm 0.23$ \\ 
            
            090328  & 0.736  & n.a.                                   & n.a.                                 & $141^{+15}_{-11}$ & $0.189_{-0.049}^{+0.065}$ & $0.67_{-0.18}^{+0.18}$ & $-31_{-8}^{+8}$ & $0.20_{-0.04}^{+0.10}$ & 24 & $3.08 \pm 0.11$ \\ 
            
            090424  & 0.544  & $<23.1$                                & $<16.1$                              & $300^{+79}_{-79}$ & $0.073_{-0.019}^{+0.025}$ & $0.44_{-0.06}^{+0.06}$ & $-21_{-3}^{+3}$ & $0.09_{-0.01}^{+0.04}$ & 34 & $2.27 \pm 0.03$ \\ 
            
            091208B & 1.0633 & $<14.8$                                & $<13.8$                              & $500^{+33}_{-33}$ & $0.092_{-0.024}^{+0.032}$ & $0.46_{-0.07}^{+0.07}$ & $-22_{-4}^{+4}$ &  $0.03_{-0.02}^{+0.06}$ & 13 & $2.32 \pm 0.05$ \\ 
            
            100414A & 1.368  & n.a.                                   & n.a.                                 & $262^{+8}_{-7}$   & $0.105_{-0.027}^{+0.036}$ & $0.72_{-0.52}^{+0.51}$ & $-43_{-20}^{+20}$ & $0.24_{-0.04}^{+0.17}$ & 16 & $2.67 \pm 0.31$ \\ 
            
            100728A & 1.567  & n.a.                                   & n.a.                                 & $250^{+9}_{-8}$   & $0.101_{-0.026}^{+0.035}$ & $0.45_{-0.10}^{+0.10}$ & $-21_{-4}^{+4}$ & $0.12_{-0.02}^{+0.04}$ & 46 & $2.77 \pm 0.03$ \\ 
            
            100906A & 1.727  & $2.9 \pm 0.1$                         & $3.0 \pm 0.1$                         & $<369$            & $0.917_{-0.236}^{+0.317}$ & $0.70_{-0.16}^{+0.17}$ & $-26_{-5}^{+5}$ & $0.07_{-0.03}^{+0.15}$ & 9  & $2.32 \pm 0.08$ \\ 
            
            120711A & 1.405  & n.a.                                   & n.a.                                 & $250^{+7}_{-7}$   & $0.083_{-0.021}^{+0.029}$ & $0.65_{-0.15}^{+0.14}$ & $-43_{-6}^{+6}$ & $0.15_{-0.03}^{+0.08}$ & 27 & $2.70 \pm 0.09$ \\ 
            
            140508A & 1.027  & n.a.                                   & n.a.                                 & $500^{+17}_{-16}$ & $1.044_{-0.268}^{+0.361}$ & $0.63_{-0.11}^{+0.11}$ & $-22_{-5}^{+5}$ & $0.21_{-0.03}^{+0.08}$ & 32 & $2.51 \pm 0.08$ \\ 
            
            140512A & 0.725  & $4.7 \pm 0.1$                          & $5.2 \pm 0.1$                        & $92^{+11}_{-8}$   & $0.434_{-0.112}^{+0.150}$ & $0.47_{-0.12}^{+0.12}$ & $-28_{-3}^{+3}$ &  $0.10_{-0.02}^{+0.07}$ & 21 & $2.99 \pm 0.13$ \\
            
            151027A & 0.81   & $7.9 \pm 0.9$                          & $8.3 \pm 0.6$                        & $143^{+20}_{-17}$ & $1.845_{-0.474}^{+0.638}$ & $0.57_{-0.28}^{+0.29}$ & $-21_{-3}^{+3}$ & $0.23_{-0.04}^{+0.18}$ & 15 & $2.83 \pm 0.33$ \\
            
            160509A & 1.17   & $9.2 \pm 0.7$                          & $5.4 \pm 0.3$                        & $470^{+22}_{-21}$ & $0.515_{-0.132}^{+0.178}$ & $0.27_{-0.06}^{+0.06}$ & $-12_{-3}^{+3}$ & $0.15_{-0.02}^{+0.04}$ & 50 & $2.63 \pm 0.04$ \\
            
            160625B & 1.406  & $14.8 \pm 1.0$                         & $6.0 \pm 0.3$                        & $860^{+19}_{-19}$ & $0.174_{-0.045}^{+0.060}$ & $0.39_{-0.06}^{+0.06}$ & $-18_{-3}^{+3}$ & $0.16_{-0.02}^{+0.06}$ & 73 & $2.52 \pm 0.07$ \\
            
            170405A & 3.510  & $2.0 \pm 0.1$                          & $2.8 \pm 0.1$                        & n.a.              & $0.605_{-0.155}^{+0.209}$ & $0.32_{-0.17}^{+0.18}$ & $-17_{-5}^{+5}$ & $0.12_{-0.03}^{+0.07}$ & 23 & $2.78 \pm 0.17$ \\
            
            171010A$^{\rm (a)}$ & 0.3285 & $<11.0$                    & n.a.                                 & $201^{+7}_{-7}$   & $0.364_{-0.094}^{+0.126}$ & $0.39_{-0.05}^{+0.05}$ & $-22_{-3}^{+3}$ & $0.19_{-0.02}^{+0.02}$ & 162 & $2.46 \pm 0.05$ \\
            
            180720B & 0.654  & $7.2 \pm 0.7$                         & $5.3 \pm 0.3$                         & n.a.              & $0.090_{-0.023}^{+0.031}$ & $0.36_{-0.07}^{+0.07}$ & $-17_{-3}^{+3}$ & $0.19_{-0.02}^{+0.03}$ & 93 & $2.66 \pm 0.03$ \\
            
            181020A & 2.938  & $2.1 \pm 0.3$                         & $2.0 \pm 0.2$                         & n.a.              & $0.376_{-0.097}^{+0.130}$ & $0.39_{-0.20}^{+0.20}$ & $-16_{-2}^{+2}$ & $0.09_{-0.02}^{+0.11}$ & 10 & $2.65 \pm 0.22$ \\
            
            190114C & 0.4245 & $3.2 \pm 0.5$                         & $3.2 \pm 0.3$                         & n.a.              & $0.098_{-0.025}^{+0.034}$ & $0.65_{-0.06}^{+0.06}$ & $-32_{-2}^{+2}$ & $0.23_{-0.03}^{+0.05}$ & 64 & $2.55 \pm 0.04$ \\
            
            221010A$^{\rm (b)}$ & 4.615  & $6_{-2}^{+1}$             & n.a.                                  & n.a.              & n.a.                      & $0.40_{-0.10}^{+0.10}$ & $-20_{-2}^{+2}$ & $0.11_{-0.02}^{+0.05}$ & 31 & $2.55 \pm 0.10$ \\
            \hline
            \end{tabular}
            \begin{list}{}{}
                \item [$^{\rm (a)}$] $\theta_{\rm j}^{\rm (ISM)}$ taken from \citet{Chand19}. 
                \item [$^{\rm (b)}$] $z$ and $\theta_{\rm j}^{\rm (ISM)}$ taken from \citet{Zhu23}.
            \end{list}
        \end{table*}
        
    \subsection{Time-resolved spectral analysis}
        \label{subsec::spectralanalysis}
        
        The time-resolved spectral analysis of each burst in the sample $\mathcal{S}$ was carried out with the GBM Data Tools\footnote{\url{https://fermi.gsfc.nasa.gov/ssc/data/analysis/gbm/}} as follows.
        
        We used the TTE data of the two most illuminated NaI detectors. LCs were binned with a common bin time $\Delta t$, which ranged from a minimum of 64~ms to a maximum of 1024~ms according to the formula $\Delta t = 64 \times 2^n$~ms, with $n = 0, \ldots, 4$. The bin time was chosen so as to achieve the highest S/N, with a maximum of 1024~ms to preserve as much as possible the genuine structures in the LCs. Then, we summed them to further increase the S/N. This required a preliminary background subtraction from both of the LCs, which had been done by interpolating the background with a polynomial function up to third order. The total LC, which we obtained from the procedure described above, was the starting point of our work. 
        
        Time intervals were determined adopting the scheme given by Bayesian Blocks (BBs; \citealt{Scargle13}) as implemented in {\tt astropy.stats} Python library. The choice of BBs was driven by the need to resolve spectrally different temporal structures in the burst LC in an unbiased way \citep{Burgess14}. At the same time, to ensure enough statistics, we adopted the following rule of thumb: $\ge 1000$~counts had to be contained in each final interval, starting from the scheme given by BBs. Consequently, BBs were either taken as standalone intervals, or grouped so as to match our requirement on the minimum number of counts. The number of intervals, $N$, in which the LC of each GRB has been split is reported in Table~\ref{table::intrinsicdispersion}.
        
        Energy spectra in each selected final interval were fitted with a Band function \citep{Band93}, also to ensure homogeneity, where we fixed $\beta = -2.3$, since in most cases the high-energy tail of the 8--1000~keV spectrum was poorly constrained (see also the Fermi spectral catalogue by \citealt{Poolakkil21}). We were therefore able to constrain: the low-energy index, $\alpha$, the peak energy, $E_{\rm p}$, and the average energy flux, $F$. Uncertainties on $\alpha$ and $E_{\rm p}$ were computed with GBM tools. For what concerns $F$, uncertainties were computed as follows. We approximated the posterior probability distribution function (PDF) of the model parameters around its maximum with a multi-dimensional normal distribution, whose mean and covariance matrix corresponded to the model parameter vector and covariance matrix derived from the fit, respectively. We then generated a random sample of model parameter vectors, $V_i$ with $i = 1, \ldots, 1000$, drawn from the multi-dimensional normal distribution. For each $V_i$, we computed the energy flux, $F_i$, with the GBM tools. We computed the standard deviation of the $F_i$ sample and then converted it to 90\% confidence level uncertainties.

        The last step was moving from the observer-frame quantities $E_{\rm p}$ and $F$ to the corresponding intrinsic quantities, namely $E_{\rm p,i}$ and $L_{\rm iso}$, exploiting the knowledge of the redshift. The isotropic-equivalent luminosity was computed in the GRB-rest-frame standard energy band $1 - 10^4$~keV. 

        We obtained a total of 793 time-resolved spectra for $\mathcal{S}$, 421 for $\mathcal{S}^{\rm (ISM)}$, and 390 for $\mathcal{S}^{\rm (W)}$.
        
    \subsection{Modelling of the time-resolved $E_{\rm p,i}$--$L_{\rm iso}$ correlation}
        \label{subsec::timeresolvedamaticorrelation}
        
\begin{figure*}[!h]
            \centering
            \includegraphics[width=17cm]{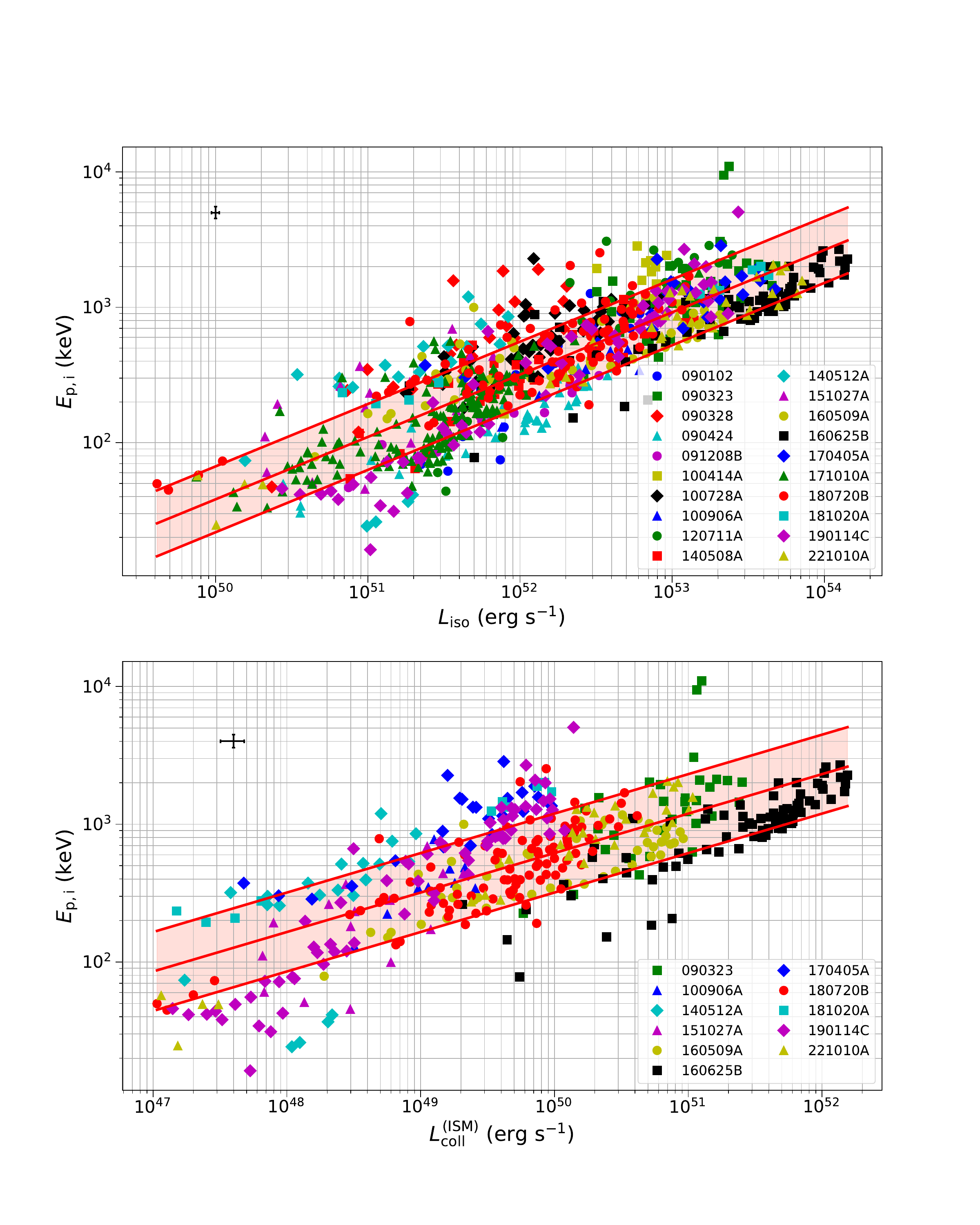}
            \caption{{\em Top panel}: time-resolved $E_{\rm p,i}$--$L_{\rm iso}$ relation for the overall sample $\mathcal{S}$. {\em Bottom panel}: collimation-corrected $E_{\rm p,i}$--$L_{\rm coll}$ relation for the subset $\mathcal{S}^{\rm (ISM)}$ of GRBs with available information on $\theta_{\rm j}^{\rm (ISM)}$. In both panels the best-fit model along with the 1-$\sigma_{\mathrm{int, t}}$ region a (shaded-area) is shown, while median uncertainties along both axes are shown in the top left.}
            \label{fig::amaticorrelation}
        \end{figure*}

        For each burst we modelled the $E_{\rm p,i}$--$L_{\rm iso}$ correlation within a Bayesian context, adopting the D'Agostini likelihood \citep{D'Agostini05}, which self-consistently treats the intrinsic dispersion of the correlation as one of the model parameters. A uniform prior was assumed for the model parameters $m$, $q$, and $\sigma_{\rm int}$. Parameter uncertainties were computed at 90\% confidence level by sampling the joint posterior PDF with a Python implementation of the affine-invariant ensemble sampler for Markov Chain Monte Carlo (MCMC; \citealt{Goodman10}), using Python package {\tt emcee} \citep{Foreman-Mackey13}. Analogously, the value of dispersion, $\sigma_{\mathrm{int, t}}$, relative to the overall sample $\mathcal{S}$, was estimated merging the time-resolved pairs ($E_{\rm p,i}$, $L_{\rm iso}$) of every burst into a unique sample and analysed as if it were a single GRB (see the top panel of Figure~\ref{fig::amaticorrelation}).

    \subsection{Simulated distributions of the intrinsic dispersion of individual bursts}
        \label{subsec::simulateddistributions}

        To investigate to which extent assembling different GRBs affects the dispersion of the time-resolved $E_{\rm p,i}$--$L_{\rm iso}$ correlation, we assume that the latter is equally dispersed over the same region for all GRBs in the sample $\mathcal{S}$ (hereafter H$_0$ hypothesis). Under this assumption, points in the $E_{\rm p,i}$--$L_{\rm iso}$ plane belonging to any given GRB should be indistinguishable from random samplings of the overall sample. We exploited this predicted property to test H$_0$.

        Let $(E_{\rm p,i}$, $L_{\rm iso})_{i, j}$ be the pair of values of the $i$-th interval of the $j$-th burst, with $i = 1, \ldots, N_j$ and $j = 1, \ldots ,20$, where $N_j$ denotes the number of intervals in which the LC of the $j$-th burst has been split. The total sample $\mathcal{S}$ includes $N_{\rm tot} = \sum_{j = 1}^{20} N_j$ observed pairs $(E_{\rm p,i}$, $L_{\rm iso})$. For each round of simulations, we shuffled $\mathcal{S}$ and split it into the 20 subsets having the same corresponding numbers of elements, $N_j$ ($j = 1, \ldots , 20$) as the real GRBs. For each shuffled subset of $N_j$ pairs of ($E_{\rm p,i}$, $L_{\rm iso}$) and corresponding uncertainties we modelled the correlation following the procedure described in Section~\ref{subsec::timeresolvedamaticorrelation}. Eventually, we stored the values of the simulated intrinsic dispersion of each burst into an array $R = (\sigma_1 , \ldots , \sigma_{20})$.

        The entire procedure described above was repeated $10^4$ times. By assembling together the $R_i$ arrays, with $i = 1, \ldots, 10^4$, we built a $10^4 \times 20$ matrix whose $j$-th column contained $10^4$ simulated values of the intrinsic dispersion of the $j$-th burst in the sample ($j = 1, \ldots, 20$). The distribution of the values of column $j$ therefore represents the PDF of $\sigma_j$ under H$_0$, hereafter denoted as $f_j(\sigma_j)$. $\sigma_j$ and $\sigma_{{\rm int},j}$ denote the random variable and the measured value from real data, respectively, for the $j$-th burst. We then determined the corresponding cumulative density function (CDF) $F_j(\sigma_j) = \int_0^{\sigma_j} f_j(x) \, \mathrm{d}x$ and calculated the probability $P_j = F_j(\sigma_{{\rm int,ul}, j}$), where $\sigma_{{\rm int,ul}, j}$ is the upper limit on $\sigma_{{\rm int}, j}$. We opted for a conservative approach, so the upper limit was taken instead of the corresponding value. $P_j$ is therefore the probability that, under H$_0$, $N_j$ points randomly sampled in the $E_{\rm p,i}$-–$L_{\rm iso}$ plane from the whole set of all GRBs are no more dispersed (within uncertainties) than the $N_j$ measured points of the $j$-th GRB.

%__________________________________________________________________

    \section{Results}
        \label{sec::results}

        Table~\ref{table::intrinsicdispersion} reports the parameters of the $E_{\rm p,i}$--$L_{\rm iso}$ correlation for individual bursts. The analogous values computed for the total sample $\mathcal{S}$ are $\sigma_{\rm int, t} = 0.24_{-0.01}^{+0.01}$, $m_{\rm t} = 0.46_{-0.02}^{+0.02}$, and $q_{\rm t} = -21.4 \pm 0.9$. The slope lies within the range of values reported by other authors on different data sets ($0.36 \pm 0.05$ for \citealt{Ghirlanda10}; $0.621 \pm 0.003$ for \citealt{Lu12}; $0.63_{-0.07}^{+0.06}$ for \citealt{Frontera12}), whereas the dispersion is instead compatible with that found by \citet{Lu12} ($\sigma_{\rm int, t} = 0.256$). The discrepancy between the values of the slope presented in the literature likely reflects the heterogeneity of approaches and data sets. 

        For both of the subsets $\mathcal{S}^{\rm (ISM)}$ and $\mathcal{S}^{\rm (W)}$ we computed the parameters $m_{\rm coll}$, $q_{\rm coll}$, and $\sigma_{\rm coll}$ of the $E_{\rm p,i}$--$L_{\rm coll}$ correlation, where $L_{\rm coll} = L_{\rm iso} (1 - \cos \theta_{\rm j})$ is the collimation-corrected luminosity (see the bottom panel of Figure~\ref{fig::amaticorrelation}). The values are reported in Table~\ref{table::amaticorrelationparameters}. Both $\sigma_{\rm coll}^{\rm (ISM)}$ and $\sigma_{\rm coll}^{\rm (W)}$ are not smaller than $\sigma_{\rm int, t}$ (computed on either $\mathcal{S}^{\rm (ISM)}$ or $\mathcal{S}^{\rm (W)}$), suggesting that the the jet opening angle cannot be considered as the main source of dispersion in the $E_{\rm p,i}$--$L_{\rm iso}$ correlation, unless one admits the possibility that the way jet opening angles are routinely estimated from the afterglow modelling is not reliable. 
    
    \subsection{Dispersion of individual bursts}
        \label{subsec::dispersionofindividualbursts}

        Almost all the dispersion values of individual bursts are smaller, or at most equal to the dispersion $\sigma_{\mathrm{int}, t}$ of the overall sample. The former are with respect to the slope and normalisation of each of them, which are generally different from those computed for the entire sample. This result apparently clashes with the hypothesis $\rm H_0$ (Section~\ref{subsec::simulateddistributions}). Yet, as long as we consider the $j$-th ($j = 1, \ldots, 20$) GRB alone, its value of $P_j$ is not small enough to reject $\rm H_0$ for that GRB. However, the simultaneous occurrence of 20 independent and relatively unlikely events demands a precise calculation, which correctly accounts for the multi-trial aspect. 
        
        To this aim, we defined the variable $X = \prod_{j = 1}^{20} p_j$, where $p_j = F_j(\sigma_j)$. Using the values of $p_j = P_j$ derived in Section~\ref{subsec::simulateddistributions} on the real sample, we obtained $X_0 = 1.50 \times 10^{-19}$.
        $X$ measures the {\em relative} probability of obtaining, under H$_0$, a sample \{$\sigma_j$\}. To establish how unlikely $X_0$ is, we have to determine its CDF, $F_X(x) = P(X \le x)$. Hence, the final significance that is looked for is $P_{\rm t} = F_X(X_0)$.
        Actually, we conveniently used $Y = - \mathrm{ln} \, X = - \sum_{j = 1}^{20} \mathrm{ln} \, p_j$. Under H$_0$ each burst can be thought as an independent realisation of the same phenomenon, so the variables $p_j$ are identically and uniformly distributed in $[0,1]$. 
        Consequently, $-\mathrm{ln}\ p_j$ is exponentially distributed. Given that $Y$ is defined as the sum of $n=20$ independent, identically and exponentially-distributed variables, its PDF, $f_Y(y)$, is a Gamma distribution:
        \begin{equation}
        f_Y(y)\ =\ \frac{y^{n-1}\,e^{-y}}{\Gamma(n)}\;,
        \label{eq:gamma}
        \end{equation}
        where $\Gamma()$ is the Gamma function.
        Because of the minus sign, it is $P(X\le X_0) = P(Y\ge Y_0)$, where $Y_0 = -\ln X_0$. Consequently, it is $P_{\rm t} = \int_{Y_0}^{+\infty} f_Y(u) \mathrm{d}u = 2.69 \times 10^{-5}$, which corresponds to a confidence level of $4.2 \, \sigma$ (Gaussian). 
        We therefore reject H$_0$ at this confidence level and conclude that GRBs are less dispersed when considered individually rather than collectively.

    \subsection{Searching for correlation between dispersion and other key observables}
        \label{subsec::searchforcorrelations}

        Motivated by the possibility that the dispersion intrinsic to the $E_{\rm p,i}$--$L_{\rm iso}$ correlation could be related to a structured, relativistic jet which wobbles around the line of sight, following the approach proposed by \citet{Camisasca23a}, we searched for correlations between $\sigma_{\rm int}$ and some key observables, namely $\theta_{\rm j}$, the $\Gamma_0$, and the MVT. For the latter, we used the same values computed by \citet{Camisasca23a}, which are reported in Table~\ref{table::intrinsicdispersion}. The only exception was 221010A, which was not in that sample. No evident correlation emerged. We also investigated whether $\sigma_{\rm int}$ is to some extent affected by the different number of spectra of each GRB, finding no evidence for it. For completeness, we repeated the same analysis replacing $\sigma_{\rm int}$ with the PL index $m$, and found nothing worth mentioning.
        
        Given that type-I GRBs are shifted upward in the $E_{\rm p,i}$--$L_{\rm iso}$ plane with respect to type-II GRBs, we investigated whether more dispersed bursts might be located closer to the former rather than the latter. To this aim, we searched for a relationship between dispersion and burst position in the intrinsic plane, where the latter was characterised as follows. For each GRB, both quantities were centred by subtracting the corresponding average values: $\log E_{\rm p,i}-\langle \log E_{\rm p,i} \rangle$, and $\log L_{\rm iso}-\langle \log L_{\rm iso} \rangle$. Using the centred values, we computed the parameters $m_{\mathrm{c}, j}$, $q_{\mathrm{c}, j}$, and $\sigma_{\mathrm{int,c}, j}$ as described in Section~\ref{subsec::timeresolvedamaticorrelation}. We then calculated a representative value for the peak energy of any given $j$-th GRB, which was evaluated at a common luminosity given by the average of the total sample, $\langle \log L_{\rm iso} \rangle_{\rm t}$, and defined as $(\log E_{\rm p,i})_{\mathrm{c}, j} = \langle \log E_{\rm p,i} \rangle_j + m_{\mathrm{c}, j} (\langle \log L_{\rm iso} \rangle_{\rm t}-\langle \log L_{\rm iso} \rangle_j) + q_{\mathrm{c}, j}$ ($E_{\rm p,i}$ and $L_{\rm iso}$ are meant to be expressed in keV and $10^{52}$~erg~s$^{-1}$, respectively). The values of $(\log E_{\rm p,i})_{\mathrm{c}}$ are reported in Table~\ref{table::intrinsicdispersion}. We found no correlation between $\sigma_{\rm int}$ and $(\log E_{\rm p,i})_{\mathrm{c}}$.

        \begin{table*}[!h]
            \centering
            \caption{Best-fit parameters of the $E_{\rm p,i}$--$L$ correlation (90\% confidence) for the overall sample $\mathcal{S}$ and for the sub-samples $\mathcal{S}^{\rm (ISM)}$ and $\mathcal{S}^{\rm (W)}$ of time-resolved spectra of GRBs with available information on $\theta_{\rm j}^{\rm (ISM)}$ and on $\theta_{\rm j}^{\rm (W)}$, respectively.  $N_{\rm int}$ and $N_{\rm GRB}$ are the total number of intervals and the number of GRBs, respectively, while $\bar{m} \pm \sigma_m$, $\bar{q} \pm \sigma_q$, and $\bar{\sigma}_{\rm int} \pm \sigma_{\sigma}$ are the mean values and standard deviations of the distributions of the best-fit parameters of individual GRBs reported in Table~\ref{table::intrinsicdispersion}. Luminosity is either isotropic-equivalent (iso) or collimation-corrected (coll).}
            \label{table::amaticorrelationparameters}
            \begin{tabular}{l c c c c c c}
            \hline\hline
                                                        & Set                           & $N_{\rm int}$ & $N_{\rm GRB}$ & $m_{\rm t}$               & $q_{\rm t}$                   & $\sigma_{\rm int,t}$ \\
            \hline
            $E_{\rm p,i}$--$L_{\rm iso}$                & $\mathcal{S}$                 & 793           & 20            & $0.46_{-0.02}^{+0.02}$    & $-21.43_{-0.85}^{+0.88}$      & $0.24_{-0.01}^{+0.01}$ \\
            $E_{\rm p,i}$--$L_{\rm iso}$                & $\mathcal{S}^{\rm (ISM)}$     & 421           & 11            & $0.41_{-0.02}^{+0.02}$    & $-18.60_{-1.06}^{+1.09}$      & $0.23_{-0.01}^{+0.01}$ \\
            $E_{\rm p,i}$--$L_{\rm iso}$                & $\mathcal{S}^{\rm (W)}$       & 390           & 10            & $0.41_{-0.02}^{+0.02}$    & $-18.53_{-1.18}^{+1.15}$      & $0.24_{-0.01}^{+0.02}$ \\
            \hline\hline
                                                        &                               &               &               & $\bar{m} \pm \sigma_m$       & $\bar{q} \pm \sigma_q$           & $\bar{\sigma}_{\rm int} \pm \sigma_\sigma$ \\
            \hline
            $E_{\rm p,i}$--$L_{\rm iso}$                & $\mathcal{S}$                 & 793           & 20            & $0.53\pm 0.17$    & $-25\pm 9$               & $0.19\pm 0.09$ \\
            \hline\hline
                                                        &                               &               &               & $m_{\rm coll}$            & $q_{\rm coll}$                & $\sigma_{\rm coll}$ \\
            \hline
            $E_{\rm p,i}$--$L_{\rm coll}^{\rm (ISM)}$   & $\mathcal{S}^{\rm (ISM)}$     & 421           & 11            & $0.29_{-0.02}^{+0.02}$    & $-11.53_{-1.02}^{+1.01}$      & $0.29_{-0.02}^{+0.02}$ \\
            $E_{\rm p,i}$--$L_{\rm coll}^{\rm (W)}$     & $\mathcal{S}^{\rm (W)}$       & 390           & 10            & $0.38_{-0.02}^{+0.02}$    & $-15.97_{-1.15}^{+1.16}$      & $0.25_{-0.02}^{+0.02}$ \\
            \hline
            \end{tabular}
        \end{table*}

    \subsection {On the robustness of the results}

        As discussed by \citet{Burgess14}, the BB method alone does not ensure enough statistical quality for spectral modelling in each interval, as is instead the case when the source is more intense than the underlying background. In our analysis this was not always the case for bursts with low S/N. Therefore, we tested the impact of our choice on the correlation parameters, as follows. We considered 090102, which had the lowest S/N that still allowed us to match the above prescription (i.e. source counts $>$ background counts). We grouped different adjacent BBs, at the same time making sure that there were enough intervals and corresponding points in the $E_{\rm p,i}$--$L_{\mathrm{iso}}$ plane to compute the correlation parameters as in Section~\ref{subsec::timeresolvedamaticorrelation}. We obtained the following values: $m_{\mathrm{B}} = 0.9 \pm 0.2$, $q_{\mathrm{B}} = -44 \pm 12$, and $\sigma_{\mathrm{int, B}} = 0.14_{-0.07}^{+0.18}$, which are compatible within uncertainties with those in Table~\ref{table::intrinsicdispersion}. This suggests that our choice of $>1000$ counts in each selected interval does not affect the correlation modelling even when the above prescription of a dominant source over background is not fulfilled.  Other GRBs with better average S/N for the different time intervals are less affected by this potential bias, whose impact can be therefore neglected a fortiori.
        \begin{table}[!h]
            \centering
            \caption{Best-fit parameters of the $E_{\rm p,i}$--$L_{\rm iso}$ correlation (90\% confidence) of the GRBs with the highest S/N obtained including BGO data with better viewing angle ($m_{\rm BGO}$, $q_{\rm BGO}$, $\sigma_{\rm int, BGO}$ using Band function; $m_{\rm CPL}$, $q_{\rm CPL}$, $\sigma_{\rm int, CPL}$ using CPL model) and NaI alone ($m$, $q$, $\sigma_{\rm int}$). $N_{\rm BGO}$ and $N_{\beta}$ are the number of selected intervals and those with estimated $\beta$ including BGO data, respectively.}
            \begin{tabular}{l c c c c}
            \hline\hline
            GRB     & $m_{\rm BGO}$     & $q_{\rm BGO}$ & $\sigma_{\rm int, BGO}$   & $N_{\beta}/N_{\rm BGO}$   \\
            \hline
            160625B & $0.36 \pm 0.09$   & $-16 \pm 5$   & $0.19_{-0.03}^{+0.04}$    & $33/72$                   \\
            171010A & $0.48 \pm 0.06$   & $-23 \pm 3$   & $0.18_{-0.02}^{+0.02}$    & $64/159$                  \\
            180720A & $0.41 \pm 0.07$   & $-19 \pm 4$   & $0.17_{-0.02}^{+0.03}$    & $10/86$                   \\
            \hline\hline
            GRB     & $m_{\rm CPL}$     & $q_{\rm CPL}$ & $\sigma_{\rm int, CPL}$   &                           \\
            \hline
            160625B & $0.44 \pm 0.08$   & $-20 \pm 4$   & $0.20_{-0.03}^{+0.04}$    &                           \\
            171010A & $0.47 \pm 0.05$   & $-22 \pm 3$   & $0.17_{-0.02}^{+0.02}$    &                           \\
            180720A & $0.46 \pm 0.07$   & $-21 \pm 4$   & $0.17_{-0.02}^{+0.03}$    &                           \\
            \hline\hline
            GRB     & $m$               & $q$           & $\sigma_{\rm int}$        &                           \\
            \hline
            160625B & $0.39 \pm 0.06$   & $-18 \pm 3$   & $0.16_{-0.02}^{+0.06}$    &                           \\
            171010A & $0.39 \pm 0.05$   & $-22 \pm 3$   & $0.19_{-0.02}^{+0.02}$    &                           \\
            180720A & $0.36 \pm 0.07$   & $-17 \pm 3$   & $0.19_{-0.02}^{+0.03}$    &                           \\
            \hline
            \end{tabular}
            \label{tab::correlation_parameters_bgo}
        \end{table}      

        With reference to Section~\ref{subsec::spectralanalysis}, we explored the alternative approach of constraining the high-energy index of the Band function, by also using the data from the BGO detector with the more favourable viewing angle. We evaluated the impact on the correlation parameters of the three GRBs with the highest S/N and a high BGO signal, which are 160625B, 171010A, and 180720A. Whenever $\beta$ could not be constrained through spectral analysis, we kept fixing it to $-2.3$. Results are reported in Table~\ref{tab::correlation_parameters_bgo}. We called $N_{\rm BGO}$ and $N_{\beta}$ the number of selected intervals and the number of intervals with estimated $\beta$, respectively. All parameters are compatible within uncertainties, suggesting that having neglected BGO data does not impact appreciably on the modelling of the $E_{\rm p,i}$--$L_{\mathrm{iso}}$ correlation. In addition, all distributions of estimated $\beta$ peak around $-2.3$, which justifies our approach.

        Given that only a fraction of GRBs is preferentially fitted by the Band function, following Sections~\ref{subsec::spectralanalysis} and \ref{subsec::timeresolvedamaticorrelation} we replicated the analysis (inclusive of BGO data) choosing the alternative cutoff power law (CPL), and investigated the impact on the parameters of GRB~160625B, 171010A, and 180720A, for the same reason mentioned above. Results are presented in Table~\ref{tab::correlation_parameters_bgo}. Even in this case, all parameters are compatible with each other. We thus conclude that the model choice, such as Band vs. CPL, does not seem to impact the correlation modelling significantly. The goodness of the best-fit modelling was computed for every interval with the PGSTAT (Profile Gaussian likelihood) statistics with GBM tools, according to which neither the Band function nor the CPL model could be rejected.

%__________________________________________________________________

    \section{Discussion and conclusions}
        \label{sec::conclusions} 
        Through a time-resolved spectral analysis of a sample of 20 type-II GRBs detected by Fermi-GBM with available estimates of $z$, $\theta_{\rm j}$, and/or $\Gamma_0$, we proved that the intrinsic dispersion of the $E_{\rm p,i}$--$L_{\rm iso}$ correlation within individual GRBs is smaller than that of the sample including all of them ($4.2\sigma$ confidence). This result is not an artefact of the different number of time-resolved intervals considered and propels the dispersion of the correlation as an intrinsic property of each GRB, whose origin is to be understood yet. It also confirms that the origin of the dispersion in the sibling time-average correlations, $E_{\rm p,i}$--$E_{\rm iso}$ and $E_{\rm p,i}$--$L_{\rm p,iso}$, is to be partially ascribed to assembling together different sources. 

        We searched for correlations between $\sigma_{\rm int}$ and some key properties, such as the jet half-opening angle, the Lorentz factor, and the MVT, that could bring signatures of a randomly wobbling jet, as suggested by recent state-of-the-art simulations, finding no evidence. 

        When the time-resolved isotropic-equivalent luminosity is replaced by the corresponding collimation-corrected estimates, either assuming an ISM or a wind environment, the dispersion of the $E_{\rm p,i}$--$L_{\rm coll}$ correlation does not decrease at all, proving that the jet opening angle cannot be considered as a significant source of dispersion, unless the estimates for the angles obtained in the literature from the afterglow modelling are, for some unknown reasons, unreliable. Instead, the viewing angle could be a relevant source of scattering, given that the correlation itself could also be interpreted as a viewing angle effect for a variety of jet models (see \citealt{Salafia15} and references therein). However, estimating the viewing angle from afterglow modelling is presently more challenging.
        
        We also investigated whether bursts lying closer to type-I GRBs in the $E_{\rm p,i}$--$L_{\rm iso}$ plane, therefore potentially misidentified compact binary merger candidates, exhibit a significantly different dispersion from the bulk of collapsar candidates, but no evidence for it was found. Yet, our search was significantly hampered by the low statistical sensitivity due to the small number of GRBs with the available information. Only a larger sample of GRBs with estimates of the properties mentioned above will allow us to explore with the required sensitivity. In this respect, a major breakthrough will be achieved thanks to upcoming missions like SVOM \citep{Atteia22_SVOM}, possibly followed by THESEUS \citep{Amati21}, which has recently been selected by ESA for a Phase-A study within the M7 mission context. These missions will cover the entire spectrum of the prompt emission, thus constraining the spectral parameters of soft events and extending the dynamic range in the $E_{\rm p,i}$--$L_{\rm iso}$ plane. In parallel, the followup segments will contribute to increase the sample of GRBs with estimates of the above key properties. Ultimately, this will turn into a systematic and more accurate characterisation of the $E_{\rm p,i}$--$L_{\rm iso}$ correlation, that will constrain the dissipation mechanism and the radiative processes that shape GRB prompt emission.
        
%__________________________________________________________________

\begin{acknowledgements}
The constructive feedback from the anonymous reviewer is gratefully acknowledged. M.M. and R.M. acknowledge the University of Ferrara for the financial support of their PhD scholarships.
% Froh zu sein bedarf es wenig, und wer froh ist, ist ein König.
\end{acknowledgements}

    \bibpunct{(}{)}{;}{a}{}{,}
    \bibliographystyle{aa}
    \bibliography{refs}

\begin{thebibliography}{33}
\expandafter\ifx\csname natexlab\endcsname\relax\def\natexlab#1{#1}\fi

\bibitem[{{Amati} {et~al.}(2002){Amati}, {Frontera}, {Tavani}, {in't Zand}, {Antonelli}, {Costa}, {Feroci}, {Guidorzi}, {Heise}, {Masetti}, {Montanari}, {Nicastro}, {Palazzi}, {Pian}, {Piro}, \& {Soffitta}}]{Amati02}
{Amati}, L., {Frontera}, F., {Tavani}, M., {et~al.} 2002, \aap, 390, 81

\bibitem[{{Amati} {et~al.}(2021){Amati}, {O'Brien}, {G{\"o}tz}, {Bozzo}, {Santangelo}, {Tanvir}, {Frontera}, {Mereghetti}, {Osborne}, {Blain}, {Basa}, {Branchesi}, {Burderi}, {Caballero-Garc{\'\i}a}, {Castro-Tirado}, {Christensen}, {Ciolfi}, {De Rosa}, {Doroshenko}, {Ferrara}, {Ghirlanda}, {Hanlon}, {Heddermann}, {Hutchinson}, {Labanti}, {Le Floch}, {Lerman}, {Paltani}, {Reglero}, {Rezzolla}, {Rosati}, {Salvaterra}, {Stratta}, {Tenzer}, \& {Theseus Consortium}}]{Amati21}
{Amati}, L., {O'Brien}, P.~T., {G{\"o}tz}, D., {et~al.} 2021, Experimental Astronomy, 52, 183

\bibitem[{{Atteia} {et~al.}(2022){Atteia}, {Cordier}, \& {Wei}}]{Atteia22_SVOM}
{Atteia}, J.~L., {Cordier}, B., \& {Wei}, J. 2022, International Journal of Modern Physics D, 31, 2230008

\bibitem[{{Band} {et~al.}(1993){Band}, {Matteson}, {Ford}, {Schaefer}, {Palmer}, {Teegarden}, {Cline}, {Briggs}, {Paciesas}, {Pendleton}, {Fishman}, {Kouveliotou}, {Meegan}, {Wilson}, \& {Lestrade}}]{Band93}
{Band}, D., {Matteson}, J., {Ford}, L., {et~al.} 1993, \apj, 413, 281

\bibitem[{{Basak} \& {Rao}(2013)}]{Basak13}
{Basak}, R. \& {Rao}, A.~R. 2013, \mnras, 436, 3082

\bibitem[{{Burgess}(2014)}]{Burgess14}
{Burgess}, J.~M. 2014, \mnras, 445, 2589

\bibitem[{{Camisasca} {et~al.}(2023{\natexlab{a}}){Camisasca}, {Guidorzi}, {Amati}, {Frontera}, {Song}, {Xiao}, {Xiong}, {Zhang}, {Margutti}, {Kobayashi}, {Mundell}, {Ge}, {Gomboc}, {Jia}, {Jordana-Mitjans}, {Li}, {Li}, {Maccary}, {Shrestha}, {Xue}, \& {Zhang}}]{Camisasca23a}
{Camisasca}, A.~E., {Guidorzi}, C., {Amati}, L., {et~al.} 2023{\natexlab{a}}, \aap, 671, A112

\bibitem[{{Camisasca} {et~al.}(2023{\natexlab{b}}){Camisasca}, {Guidorzi}, {Bulla}, {Amati}, {Rossi}, {Stratta}, \& {Singh}}]{Camisasca23b}
{Camisasca}, A.~E., {Guidorzi}, C., {Bulla}, M., {et~al.} 2023{\natexlab{b}}, GRB Coordinates Network, 33577, 1

\bibitem[{{Chand} {et~al.}(2019){Chand}, {Chattopadhyay}, {Oganesyan}, {Rao}, {Vadawale}, {Bhattacharya}, {Bhalerao}, \& {Misra}}]{Chand19}
{Chand}, V., {Chattopadhyay}, T., {Oganesyan}, G., {et~al.} 2019, \apj, 874, 70

\bibitem[{{D'Agostini}(2005)}]{D'Agostini05}
{D'Agostini}, G. 2005, arXiv e-prints, physics/0511182

\bibitem[{{Foreman-Mackey} {et~al.}(2013){Foreman-Mackey}, {Hogg}, {Lang}, \& {Goodman}}]{Foreman-Mackey13}
{Foreman-Mackey}, D., {Hogg}, D.~W., {Lang}, D., \& {Goodman}, J. 2013, \pasp, 125, 306

\bibitem[{{Frontera} {et~al.}(2012){Frontera}, {Amati}, {Guidorzi}, {Landi}, \& {in't Zand}}]{Frontera12}
{Frontera}, F., {Amati}, L., {Guidorzi}, C., {Landi}, R., \& {in't Zand}, J. 2012, \apj, 754, 138

\bibitem[{{Ghirlanda} {et~al.}(2004){Ghirlanda}, {Ghisellini}, \& {Lazzati}}]{Ghirlanda04}
{Ghirlanda}, G., {Ghisellini}, G., \& {Lazzati}, D. 2004, \apj, 616, 331

\bibitem[{{Ghirlanda} {et~al.}(2018){Ghirlanda}, {Nappo}, {Ghisellini}, {Melandri}, {Marcarini}, {Nava}, {Salafia}, {Campana}, \& {Salvaterra}}]{Ghirlanda18}
{Ghirlanda}, G., {Nappo}, F., {Ghisellini}, G., {et~al.} 2018, \aap, 609, A112

\bibitem[{{Ghirlanda} {et~al.}(2010){Ghirlanda}, {Nava}, \& {Ghisellini}}]{Ghirlanda10}
{Ghirlanda}, G., {Nava}, L., \& {Ghisellini}, G. 2010, \aap, 511, A43

\bibitem[{{Goodman} \& {Weare}(2010)}]{Goodman10}
{Goodman}, J. \& {Weare}, J. 2010, Communications in Applied Mathematics and Computational Science, 5, 65

\bibitem[{{Gottlieb} {et~al.}(2022){Gottlieb}, {Liska}, {Tchekhovskoy}, {Bromberg}, {Lalakos}, {Giannios}, \& {M{\"o}sta}}]{Gottlieb22}
{Gottlieb}, O., {Liska}, M., {Tchekhovskoy}, A., {et~al.} 2022, \apjl, 933, L9

\bibitem[{{Kumar} \& {Zhang}(2015)}]{Kumar15}
{Kumar}, P. \& {Zhang}, B. 2015, \physrep, 561, 1

\bibitem[{{Lu} {et~al.}(2012){Lu}, {Wei}, {Liang}, {Zhang}, {L{\"u}}, {L{\"u}}, {Lei}, \& {Zhang}}]{Lu12}
{Lu}, R.-J., {Wei}, J.-J., {Liang}, E.-W., {et~al.} 2012, \apj, 756, 112

\bibitem[{{Meegan} {et~al.}(2009){Meegan}, {Lichti}, {Bhat}, {Bissaldi}, {Briggs}, {Connaughton}, {Diehl}, {Fishman}, {Greiner}, {Hoover}, {van der Horst}, {von Kienlin}, {Kippen}, {Kouveliotou}, {McBreen}, {Paciesas}, {Preece}, {Steinle}, {Wallace}, {Wilson}, \& {Wilson-Hodge}}]{Meegan09}
{Meegan}, C., {Lichti}, G., {Bhat}, P.~N., {et~al.} 2009, \apj, 702, 791

\bibitem[{{Minaev} \& {Pozanenko}(2020)}]{MinaevPozanenko20}
{Minaev}, P.~Y. \& {Pozanenko}, A.~S. 2020, \mnras, 492, 1919

\bibitem[{{Moresco} {et~al.}(2022){Moresco}, {Amati}, {Amendola}, {Birrer}, {Blakeslee}, {Cantiello}, {Cimatti}, {Darling}, {Della Valle}, {Fishbach}, {Grillo}, {Hamaus}, {Holz}, {Izzo}, {Jimenez}, {Lusso}, {Meneghetti}, {Piedipalumbo}, {Pisani}, {Pourtsidou}, {Pozzetti}, {Quartin}, {Risaliti}, {Rosati}, \& {Verde}}]{Moresco22}
{Moresco}, M., {Amati}, L., {Amendola}, L., {et~al.} 2022, Living Reviews in Relativity, 25, 6

\bibitem[{{Planck Collaboration} {et~al.}(2020){Planck Collaboration}, {Aghanim}, {Akrami}, {Ashdown}, {Aumont}, {Baccigalupi}, {Ballardini}, {Banday}, {Barreiro}, {Bartolo}, {Basak}, {Battye}, {Benabed}, {Bernard}, {Bersanelli}, {Bielewicz}, {Bock}, {Bond}, {Borrill}, {Bouchet}, {Boulanger}, {Bucher}, {Burigana}, {Butler}, {Calabrese}, {Cardoso}, {Carron}, {Challinor}, {Chiang}, {Chluba}, {Colombo}, {Combet}, {Contreras}, {Crill}, {Cuttaia}, {de Bernardis}, {de Zotti}, {Delabrouille}, {Delouis}, {Di Valentino}, {Diego}, {Dor{\'e}}, {Douspis}, {Ducout}, {Dupac}, {Dusini}, {Efstathiou}, {Elsner}, {En{\ss}lin}, {Eriksen}, {Fantaye}, {Farhang}, {Fergusson}, {Fernandez-Cobos}, {Finelli}, {Forastieri}, {Frailis}, {Fraisse}, {Franceschi}, {Frolov}, {Galeotta}, {Galli}, {Ganga}, {G{\'e}nova-Santos}, {Gerbino}, {Ghosh}, {Gonz{\'a}lez-Nuevo}, {G{\'o}rski}, {Gratton}, {Gruppuso}, {Gudmundsson}, {Hamann}, {Handley}, {Hansen}, {Herranz}, {Hildebrandt}, {Hivon}, {Huang}, {Jaffe}, {Jones}, {Karakci}, {Keih{\"a}nen},
  {Keskitalo}, {Kiiveri}, {Kim}, {Kisner}, {Knox}, {Krachmalnicoff}, {Kunz}, {Kurki-Suonio}, {Lagache}, {Lamarre}, {Lasenby}, {Lattanzi}, {Lawrence}, {Le Jeune}, {Lemos}, {Lesgourgues}, {Levrier}, {Lewis}, {Liguori}, {Lilje}, {Lilley}, {Lindholm}, {L{\'o}pez-Caniego}, {Lubin}, {Ma}, {Mac{\'\i}as-P{\'e}rez}, {Maggio}, {Maino}, {Mandolesi}, {Mangilli}, {Marcos-Caballero}, {Maris}, {Martin}, {Martinelli}, {Mart{\'\i}nez-Gonz{\'a}lez}, {Matarrese}, {Mauri}, {McEwen}, {Meinhold}, {Melchiorri}, {Mennella}, {Migliaccio}, {Millea}, {Mitra}, {Miville-Desch{\^e}nes}, {Molinari}, {Montier}, {Morgante}, {Moss}, {Natoli}, {N{\o}rgaard-Nielsen}, {Pagano}, {Paoletti}, {Partridge}, {Patanchon}, {Peiris}, {Perrotta}, {Pettorino}, {Piacentini}, {Polastri}, {Polenta}, {Puget}, {Rachen}, {Reinecke}, {Remazeilles}, {Renzi}, {Rocha}, {Rosset}, {Roudier}, {Rubi{\~n}o-Mart{\'\i}n}, {Ruiz-Granados}, {Salvati}, {Sandri}, {Savelainen}, {Scott}, {Shellard}, {Sirignano}, {Sirri}, {Spencer}, {Sunyaev}, {Suur-Uski}, {Tauber}, {Tavagnacco},
  {Tenti}, {Toffolatti}, {Tomasi}, {Trombetti}, {Valenziano}, {Valiviita}, {Van Tent}, {Vibert}, {Vielva}, {Villa}, {Vittorio}, {Wandelt}, {Wehus}, {White}, {White}, {Zacchei}, \& {Zonca}}]{Planck20}
{Planck Collaboration}, {Aghanim}, N., {Akrami}, Y., {et~al.} 2020, \aap, 641, A6

\bibitem[{{Poolakkil} {et~al.}(2021){Poolakkil}, {Preece}, {Fletcher}, {Goldstein}, {Bhat}, {Bissaldi}, {Briggs}, {Burns}, {Cleveland}, {Giles}, {Hui}, {Kocevski}, {Lesage}, {Mailyan}, {Malacaria}, {Paciesas}, {Roberts}, {Veres}, {von Kienlin}, \& {Wilson-Hodge}}]{Poolakkil21}
{Poolakkil}, S., {Preece}, R., {Fletcher}, C., {et~al.} 2021, \apj, 913, 60

\bibitem[{{Rastinejad} {et~al.}(2022){Rastinejad}, {Gompertz}, {Levan}, {Fong}, {Nicholl}, {Lamb}, {Malesani}, {Nugent}, {Oates}, {Tanvir}, {de Ugarte Postigo}, {Kilpatrick}, {Moore}, {Metzger}, {Ravasio}, {Rossi}, {Schroeder}, {Jencson}, {Sand}, {Smith}, {Ag{\"u}{\'\i} Fern{\'a}ndez}, {Berger}, {Blanchard}, {Chornock}, {Cobb}, {De Pasquale}, {Fynbo}, {Izzo}, {Kann}, {Laskar}, {Marini}, {Paterson}, {Escorial}, {Sears}, \& {Th{\"o}ne}}]{Rastinejad22}
{Rastinejad}, J.~C., {Gompertz}, B.~P., {Levan}, A.~J., {et~al.} 2022, \nat, 612, 223

\bibitem[{{Rossi} {et~al.}(2022){Rossi}, {Rothberg}, {Palazzi}, {Kann}, {D'Avanzo}, {Amati}, {Klose}, {Perego}, {Pian}, {Guidorzi}, {Pozanenko}, {Savaglio}, {Stratta}, {Agapito}, {Covino}, {Cusano}, {D'Elia}, {De Pasquale}, {Della Valle}, {Kuhn}, {Izzo}, {Loffredo}, {Masetti}, {Melandri}, {Minaev}, {Guelbenzu}, {Paris}, {Paiano}, {Plantet}, {Rossi}, {Salvaterra}, {Schulze}, {Veillet}, \& {Volnova}}]{Rossi22}
{Rossi}, A., {Rothberg}, B., {Palazzi}, E., {et~al.} 2022, \apj, 932, 1

\bibitem[{{Salafia} {et~al.}(2015){Salafia}, {Ghisellini}, {Pescalli}, {Ghirlanda}, \& {Nappo}}]{Salafia15}
{Salafia}, O.~S., {Ghisellini}, G., {Pescalli}, A., {Ghirlanda}, G., \& {Nappo}, F. 2015, MNRAS, 450, 3549

\bibitem[{{Scargle} {et~al.}(2013){Scargle}, {Norris}, {Jackson}, \& {Chiang}}]{Scargle13}
{Scargle}, J.~D., {Norris}, J.~P., {Jackson}, B., \& {Chiang}, J. 2013, arXiv e-prints, arXiv:1304.2818

\bibitem[{{Troja} {et~al.}(2022){Troja}, {Fryer}, {O'Connor}, {Ryan}, {Dichiara}, {Kumar}, {Ito}, {Gupta}, {Wollaeger}, {Norris}, {Kawai}, {Butler}, {Aryan}, {Misra}, {Hosokawa}, {Murata}, {Niwano}, {Pandey}, {Kutyrev}, {van Eerten}, {Chase}, {Hu}, {Caballero-Garcia}, \& {Castro-Tirado}}]{Troja22}
{Troja}, E., {Fryer}, C.~L., {O'Connor}, B., {et~al.} 2022, \nat, 612, 228

\bibitem[{{Yang} {et~al.}(2022){Yang}, {Ai}, {Zhang}, {Zhang}, {Liu}, {Wang}, {Yang}, {Yin}, {Li}, \& {L{\"u}}}]{Yang22}
{Yang}, J., {Ai}, S., {Zhang}, B.-B., {et~al.} 2022, \nat, 612, 232

\bibitem[{{Yonetoku} {et~al.}(2004){Yonetoku}, {Murakami}, {Nakamura}, {Yamazaki}, {Inoue}, \& {Ioka}}]{Yonetoku04}
{Yonetoku}, D., {Murakami}, T., {Nakamura}, T., {et~al.} 2004, \apj, 609, 935

\bibitem[{{Zhao} {et~al.}(2020){Zhao}, {Zhang}, {Zhang}, {Liang}, {Luan}, {Zhou}, {Yi}, {Wang}, \& {Zhang}}]{Zhao20}
{Zhao}, W., {Zhang}, J.-C., {Zhang}, Q.-X., {et~al.} 2020, \apj, 900, 112

\bibitem[{{Zhu} {et~al.}(2023){Zhu}, {Lei}, {Malesani}, {Fu}, {Liu}, {Xu}, {D'Avanzo}, {Ag{\"u}{\'\i} Fern{\'a}ndez}, {Fynbo}, {Gao}, {Nicuesa Guelbenzu}, {Jiang}, {Kann}, {Klose}, {Liu}, {Liu}, {De Pasquale}, {de Ugarte Postigo}, {Stecklum}, {Th{\"o}ne}, {Markku Viuho}, {Zhu}, {Li}, {Gao}, {Lu}, {Xiao}, {Zou}, {Xin}, \& {Wei}}]{Zhu23}
{Zhu}, Z.-P., {Lei}, W.-H., {Malesani}, D.~B., {et~al.} 2023, \apj, 959, 118

\end{thebibliography}

\end{document}